\documentstyle[preprint,aps,epsfig,psfig]{revtex} %

\begin{document}
\draft
\title{Non-leptonic Charmless 2-Body B-Decays in the Perturbative QCD
Approach 
\footnote{Supported in part by National Natural Science 
Foundation of China
and State Commission of Science and Technology of China }}
\vspace{2cm}

\author{
Dongsheng Du${}^{1,2}$, Deshan Yang${}^{2}$  and Guohuai Zhu${}^{2}$ 
\footnote{Email: duds@hptc5.ihep.ac.cn, yangds@hptc5.ihep.ac.cn,
zhugh@hptc5.ihep.ac.cn} }
\address{${}^1$ CCAST (World Laboratory), P.O.Box 8730, Beijing
100080, China\\
${}^2$ Institute of High Energy Physics, Chinese Academy of Sciences,
 P.O.Box 918(4), Beijing 100039, China }

\maketitle

\begin{abstract}
With the generalized factorization approximation, we calculate the 
branching ratios and $CP$ asymmetries in $B$ meson decays into two 
charmless pseudoscalar mesons. We give a new estimation of the
matrix elements of $(S+P)(S-P)$ current products with the $PQCD$ method
instead of equation of motion. We find that our results are 
comparatively smaller than those in the literature.
\end{abstract}

\vspace{1.5cm}
{\bf PACS numbers 13.25.Hw 13.20.He}

\newpage

\narrowtext
\tighten

\section{Introduction}
Penguin diagrams play an important role in charmless $B$ decays and
direct $CP$ violation. They can provide not only the necessary different
loop effects of internal u and c quarks\cite{Bander}
, but also dominate the branching
ratios of many modes of charmless $B$ decays, such as $B\rightarrow \pi K$,
$B\rightarrow KK$ and $B\rightarrow K\eta^{'}$.

As we know, the standard theoretical framework of studying non-leptonic
$B$ decays is based on the effective Hamilton approach and the
factorization approximation. The effective Hamiltonian is expressed by a
sum of the products of a
series of Wilson coefficients and four-quark operators. Unfortunately, we
have many difficulties in calculating  matrix elements of the four-quark
operators directly in exclusive non-leptonic $B$ decays, such as $B$ to
two pseudoscalar meson processes. So we have to use the factorization
assumption, usually the $BSW$ model\cite{BSW}.
 Then the mesonic matrix elements are
factorized into the product of two matrix elements of single currents,
governed by decay constants and form factors. However in the $BSW$ model,
the factorization involves the contributions of Fiertz transformations
of the four-quark operators. Using the Fiertz rearrangement,
one can find that the current-current product $(S+P)(S-P)$ matrix elements
should be taken into account. The general method to deal with $(S+P)(S-P)$ 
matrix elements is to transform them into $(V-A)(V-A)$ matrix elements 
by using equation of motion. Then one can find that $(S+P)(S-P)$ matrix 
elements are very sensitive to
the masses of light quarks. But the current masses of light quarks are
not determined precisely. Obviously, it brings large uncertainty for 
estimating the $(S+P)(S-P)$ matrix elements and the branching ratios of
charmless $B$ decays. As pointed by A. Ali, G. Kramer and 
C.D. L\"{u}\cite{Ali}, varying the light quarks masses by $\pm20\%$ yields
variation of up to $\pm25\%$ in some selected decay modes(such as $B^{\pm}
\rightarrow K^{\pm}\eta^{'}$ and $\bar{B}^0\rightarrow \bar{K}^0 \eta^{'}$.)
 In this work, 
instead of using equation of motion we will try to
apply the $PQCD$ method to recalculate the ratio of the $(S+P)(S-P)$ to
(V-A)(V-A) matrix elements at leading twist approximation, which is not
sensitive to the masses of light quarks. We think that it might have less 
uncertainties than the results obtained by using the quark equation of 
motion. So, it is necessary to recalculate the branching ratios for B
meson charmless decays by using $PQCD$ method
and compare with those by using equation of motion.
On the other hand,
nonfactorizable effects in charmless $B$ decays can not be neglected. To
compensate it for this, the general approach is to replace the number of
corlors $N_c$ by a phenomenological color parameter $N_c^{eff}$. We will
discuss the differences while $N_c^{eff}$ equal 2, 3, $\infty$
respectly.

This work is organized as follows: Section 2 gives the framework
of calculation including the effective Hamiltonian and fatorization
approximation. Section 3 is devoted to the $PQCD$ method to estimate the
ratio of the $(S+P)(S-P)$ to $(V-A)(V-A)$ matrix elements. In Section 4,
we calculate the branching ratios of charmless $B$ decays into two 
pseudoscalar mesons and their $CP$ asymmetries using the method mentioned 
above. We also give some discussions of the numerical results. Section 5 
is for the concluding remarks.

\section{calculational framework}
The $\vert\Delta B\vert=1$ effective Hamiltionian is 
\begin{equation}
{\cal{H}}_{eff}= \frac{G_F}{\sqrt{2}}
 \left[ \sum_{q=u,c} v_q \left(  C_1(\mu) Q^q_1(\mu)+ C_2(\mu)Q^q_2(\mu)
  + \sum_{k=3}^{10} C_k(\mu)Q_k(\mu)  \right) \right]+h.c.,
\end{equation}
where
$v_q=V_{qb}V_{qd}^{*}$(for $b\rightarrow d$ transition) or
$v_q=V_{qb}V_{qs}^{*}$(for $b\rightarrow s$ transition)
and $C_i(\mu)$ are Wilson coefficients which have been evaluated to 
next-to-leading order approximation.
 In the Eq.(1), the four-quark operators $Q_i$ are given by
\begin{equation}
\begin{array}{l}
\begin{array}{ll}
Q^u_1= ( \bar{u}_{\alpha} b_{\beta} )_{V-A}
         ( \bar{q}_{\beta} u_{\alpha} )_{V-A}&
Q^c_1= ( \bar{c}_{\alpha} b_{\beta} )_{V-A}
         ( \bar{q}_{\beta} c_{\alpha} )_{V-A}\\
Q^u_2= ( \bar{u}_{\alpha} b_{\alpha} )_{V-A}
         ( \bar{q}_{\beta} u_{\beta} )_{V-A}&
Q^c_2= ( \bar{c}_{\alpha} b_{\alpha} )_{V-A}
         ( \bar{q}_{\beta} c_{\beta} )_{V-A}\\
Q_3= (\bar{q}_{\alpha} b_{\alpha} )_{V-A}
      \sum\limits_{q'}
     ( \bar{q}^{'}_{\beta} q^{'}_{\beta} )_{V-A}&
Q_4= (\bar{q}_{\beta} b_{\alpha} )_{V-A}
      \sum\limits_{q'}
     ( \bar{q}^{'}_{\alpha} q^{'}_{\beta} )_{V-A}\\
Q_5= (\bar{q}_{\alpha} b_{\alpha} )_{V-A}
      \sum\limits_{q'}
      ( \bar{q}^{'}_{\beta} q^{'}_{\beta} )_{V+A}&
Q_6= (\bar{q}_{\beta} b_{\alpha} )_{V-A}   
      \sum\limits_{q'}
     ( \bar{q}^{'}_{\alpha} q^{'}_{\beta} )_{V+A}\\
Q_7= \frac{3}{2} (\bar{q}_{\alpha} b_{\alpha} )_{V-A}
      \sum\limits_{q'} e_{q'}
     ( \bar{q}^{'}_{\beta} q^{'}_{\beta} )_{V+A}&
Q_8=\frac{3}{2}  (\bar{q}_{\beta} b_{\alpha} )_{V-A}
   \sum\limits_{q'} e_{q'}
    ( \bar{q}^{'}_{\alpha} q^{'}_{\beta} )_{V+A}\\
Q_9= \frac{3}{2} (\bar{q}_{\alpha} b_{\alpha} )_{V-A}
      \sum\limits_{q'} e_{q'}
    ( \bar{q}^{'}_{\beta} q^{'}_{\beta} )_{V-A}&
Q_{10}=\frac{3}{2}  (\bar{q}_{\beta} b_{\alpha} )_{V-A}
      \sum\limits_{q'} e_{q'}
     ( \bar{q}^{'}_{\alpha} q^{'}_{\beta})_{V-A}\\
\end{array} \\
      
\end{array}
\end{equation}
with $Q^q_1$ and $Q^q_2$ being the tree operators, $Q_3-Q_6$ the $QCD$
penguin operators and $Q_7-Q_{10}$ the electroweak penguin operators.
With the renormalization group method, we can evolove the 
renormalization scheme independent Wilson coefficients
$\bar{C}_i(\mu)$ from the scale $\mu=m_W$ to $\mu=5.0GeV\approx m_B$, 
which are\cite{yangmz}
\begin{equation}
\begin{array}{llll}
\bar{C}_1 = -0.313 & \bar{C}_2 = 1.150 & \bar{C} _3 = 0.017 &
\bar{C} _4 = -0.037\\ \bar{C} _5 = 0.010 & \bar{C} _6 = -0.046 & 
\bar{C} _7 = -0.001\alpha_{em} &\bar{C} _8 = 0.049\alpha_{em} \\
\bar{C} _9 = -1.321\alpha_{em} & \bar{C} _{10} = 0.267\alpha_{em}. & 
\end{array}\\
\end{equation}
So we can express the physical amplitude 
as follows 
\begin{equation}
\langle \bf{Q}^T (\mu) \cdot \bf{C}(\mu) \rangle 
 \equiv \langle \bf{Q}^T\rangle _0 \cdot\bf{C}'(\mu)  
\end{equation} 
where $\langle \bf{Q}^T \rangle_0$ denote the tree level matrix elements
and
\begin{equation}
\begin{array}{llll}
C'_1=\bar{C}_1  & C'_2=\bar{C}_2 & C'_3=\bar{C} _3-\frac{P_s}{3}  &
C'_4=\bar{C} _4+P_s \\ C'_5=\bar{C} _5 -\frac{P_s}{3} & 
C'_6=\bar{C} _6+P_s  & 
C'_7=\bar{C} _7+P_e  &C'_8=\bar{C} _8  \\
C'_9=\bar{C} _9+P_e  &C'_{10}=\bar{C} _{10},  & \\
\end{array}
\end{equation}

\begin{eqnarray}
&&P_s=\frac{\alpha_s}{8\pi} \bar{C}_2(\mu) \left( \frac{10}{9}
-G(m_q,q^2,\mu) \right)\nonumber\\ 
&&P_e=\frac{\alpha_{em}}{9\pi} \left(3 \bar{C}_1(\mu) + \bar{C}_2(\mu) \right)
\left(\frac{10}{9} - G(m_q,q^2,\mu) \right)\\
&&G(m_q,q^2,\mu)=-4\int\limits_{0}^{1}dx~x(1-x)ln\left(\frac{m_q^2-x(1-x)q^2}
{\mu^2}\right).\nonumber
\end{eqnarray}

As noted in the introduction, we have to calculate the matrix elements of
the four-quark operators by using the factorization assumption. Here we apply
the $BSW$ model\cite{BSW}. However,
nonfactorizable effects are not negligible in the process of $B$ to two light
mesons. We will use the simpliest approach to compensate it by using only one
color parameter $N_c^{eff}$, even if there is no reason why using only one 
single  parameter $N_c^{eff}$ to explain the branching ratios of all kind
of different modes. 

For illustration, we give the amplitude of $B_u^{-}\rightarrow K^{-}\eta'$ 
as an example:
\begin{eqnarray}
&~&\langle K^{-}\eta'\vert {\cal{H}}_{eff} \vert B^{-}_u \rangle \nonumber \\
& = &\frac {G_F}{\sqrt{2}} \sum \limits_{q=u,c} v_q \left\{ (a_1
\delta_{qu}
+a_3+a_9)M^{K^{-}\eta'}_{suu} \right. \nonumber\\
&& +  (a_2\delta_{qu}+2a_4
-2a_6-\frac {a_8}{2}+\frac{a_{10}}{2})M^{\eta'K^{-}}_{uus}
\nonumber\\ 
&& +  (a_4 - a_6+a_3+\frac {a_8} {2}+\frac
{a_9} {2}
-\frac {a_{10}}{2} ) M^{\eta'K^{-}}_{sss}\nonumber\\
&& + (a_2 \delta_{qc}
-a_8+a_{10})M^{\eta'K^{-}}_{ccs}\nonumber\\
 && + \left. (-2a_5-2a_7)X^{K^{-}\eta'}_{suu}+(-2a_5+a_7)
X^{\eta'K^{-}}_{sss} \right\}, 
\end{eqnarray}
where
\begin{eqnarray}
a_{2i-1} = C_{2i}^{'} + \frac{C_{2i-1}^{'}} {N_c^{eff}},~~~~~~~~~
a_{2i} = C_{2i-1}^{'} + \frac{C_{2i}^{'}}{N_c^{eff}} ,
\end{eqnarray}
and
\begin{eqnarray}
&&M_{q_1q_2q_3}^{PP'} = \langle P \vert {(\bar{q_1} q_2)}_{V-A} \vert 0 
\rangle \langle P' \vert {(\bar{q_3} b)}_{V-A} \vert B \rangle \nonumber\\
&&X_{q_1q_2q_3}^{PP'} = \langle P \vert {(\bar{q_1} q_2)}_{S+P} \vert 0 
\rangle \langle P' \vert {(\bar{q_3} b)}_{S-P} \vert B \rangle .  
\end{eqnarray}\\
We will use the following parameterization for decay constants and form factors:

\begin{eqnarray}
&&\langle 0 \vert V_{\mu}-A_{\mu} \vert P(q)\rangle = i f_P q_{\mu} \nonumber\\
&&\langle P_2(q_2)\vert V_{\mu}-A_{\mu} \vert P_1(q_1) \rangle
=F_{+}^{P_1\rightarrow P_2}(q_{-}^2)q_{+\mu}+
F_{-}^{P_1\rightarrow P_2} (q_{-}^2)q_{-\mu},  
\end{eqnarray}
where $q_{\pm}=q_1\pm q_2$, and we use the monopole dominance assumption
for the $q_{-}^2$ dependence of the form factors: 
\begin{eqnarray}
&&F_{+}^{P_1\rightarrow P_2}(q_{-}^2)\approx
\frac{F_{+}^{P_1\rightarrow P_2}(0)}{1-q_{-}^2/m_{pol}^2},\nonumber\\
&&F_{-}^{P_1\rightarrow P_2}(q_{-}^2)\approx -\frac{m_1-m_2}{m_1+m_2}
F_{+}^{P_1\rightarrow P_2}(q_{-}^2).
\end{eqnarray}
Then we can obtain
\begin{equation}
M_{q_1q_2q_3}^{PP'}=-i f_P F_{+}^{B\rightarrow P'}(m_P^2)
\frac {m_B-m_{P'}}{m_B+m_{P'}} \left[(m_B+m_{P'})^2-m_P^2\right] 
\end{equation}
\begin{equation}
X_{q_1q_2q_3}^{PP'} = \frac {m_P^2}{((m_1+m_2)(m_3-m_b))}M_{q_1q_2q_3}^{PP'}, 
\end{equation}
where Eq.(13) is derived from the equation of motion and $m_i$
presents the mass of the light quark $q_i$ respectly($i=1,2,3$). 

Calculation in this framework have been discussed in detail in some papers
\cite{Ali,fleischer} involving the branching ratios and CP asymmetries in 
Non-leptonic charmless 2-body B decays. In these papers the
uncertainties resulting from the renormalization scale dependence,
non-factorizable contributions and the input parameters ($\alpha_s$,
quark masses and form factors) have been worked out. Further penguin
effects and the strong sensitivity of the CP asymmetries to the CKM
parameters ($\rho$, $\eta$) have been discussed there. In these uncertainties,
the uncertainty of light quark masses is mainly showed in the part of 
$(S+P)(S-P)$ matrix elements. Sometimes they are dominant terms 
in some modes of charmless $B$ decays, such as 
$B_u^{-}\rightarrow K^{-}\eta'$, in which the term $X_{ssu}^{\eta'K^{-}}$
is enhanced by the factor $\frac {m_{\eta'}}{m_s}$. Then it motivates us
to give a new estimation of the matrix elements $X_{q_1q_2q_3}^{PP'}$ 
to cancel the uncertainty of the light quark masses.

\section{$PQCD$ method}
Brodsky {\it et al. }\cite{brod} has pointed out that the factorization
formula of $PQCD$ can be applied to the exclusive $B$ decays into light 
mesons for the large momentum transfers.
One can write the amplitude 
as a convolution of a hard-scattering quark-gluon amplitudes $\phi(x,Q^2)$
which describe the fractional longitudinal momentum distribution of the quark 
and antiquark in each meson. An important feature of this formalism is that,
at high momentum transfer, long-range final state interactions between the 
outgoing hardrons can be neglected. In the case of non-leptonic weak decays 
the mass squared of the heavy meson $m_H^2$ establishes the relevant
momentum 
scale $Q^2\sim m_H^2$, so that for a sufficiently massive initial state 
the decay amplitude is of the order of $\alpha_s(Q^2)$, even without including 
loop corrections to the weak hamiltonian. The dominant contribution is 
controlled by single gluon exchange.
 
We intend to apply $PQCD$ method to estimate those hadronic matrix elements 
such as $(V-A)(V-A)$ and $(S+P)(S-P)$ at the leading twist approximation.
The wave function of $B$ meson and flavor $SU(3)$ singlet or octet 
pseudoscalar mesons are taken as:
\begin{eqnarray}
\Psi_B(x) &=\frac{1}{\sqrt{2}} \frac{I_C}{\sqrt{3}} \phi_B (x)
(\slash{\hskip -2mm}p +m_B )\gamma_5 ~, \nonumber\\
\Psi_P(y) &=\frac{1}{\sqrt{2}} \frac{I_C}{\sqrt{3}} \phi_P (y)
(\slash{\hskip -2mm}q +m_P )\gamma_5 ~,
\end{eqnarray}
where $I_C$ is an identity in color space. In $QCD$, the integration of the
distribution amplitude is related to the meson decay constant 
\begin{eqnarray}
\int \phi_P (y)dy=\frac{1}{2\sqrt{6}}f_P ~,~~~~
\int \phi_B (x)dx=\frac{1}{2\sqrt{6}}f_B ~.
\end{eqnarray}
Then we can write down the amplitude of Fig.1 as
\begin{eqnarray}
\langle P \vert \bar {q}_1 \gamma_{\mu}\gamma_5 q_2\vert 0\rangle ~_{PQCD} 
&=& 3\times \frac{1}{\sqrt{2}}\frac{1}{\sqrt{3}}
\int dy\phi_P (y) Tr \left[\gamma_5~(\slash{\hskip -2mm}q +m_P )
\gamma_{\mu}\gamma_5 \right] =f_P q_{\mu}~,\nonumber\\
\langle P \vert \bar {q}_1 \gamma_5 q_2 \vert 0\rangle ~_{PQCD} &=&
3\times \frac{1}{\sqrt{2}}\frac{1}{\sqrt{3}}
\int dy\phi_P (y) Tr \left[\gamma_5~(\slash{\hskip -2mm}q +m_P )
\gamma_5 \right]=f_P m_P~.
\end{eqnarray}
In a consistent way, we can use perturbative QCD to estimate the matrix
elements like $\langle P \vert \bar q_l \gamma_{\mu}b \vert B \rangle ~$ and
$\langle P \vert \bar q_l b \vert B \rangle ~$(Fig.2, Fig.3), where
$q_l$ denotes light quark field operator and 
we have neglected the fermi motion of quarks, while the gluons in the
Fig.2,3 are hard because
\begin{eqnarray}
k^2 =(xp -(1-y)q)^2 \simeq -x(1-y)m_B^2 \sim 1 GeV^2
\end{eqnarray}
(here we using mean values $\langle y \rangle \sim \frac{1}{2},
 \langle x \rangle \sim \epsilon_B $, with $ \epsilon_B \sim 0.05 - 0.1$
 and $x^2<<1$) so, we can neglect the 
 ${\cal{O}}(x^2 m_B^2)$ term and use perturbative QCD method to calculate 
the amplitude. It turns out to be
\begin{eqnarray}
&&\langle P \vert \bar q_l \gamma_{\mu}b \vert B \rangle ~_{PQCD}=
 -\frac{2}{3}g^2
\int dxdy \phi_B (x)\phi_P (y) \\
&&\left\{
\frac{Tr \left[\gamma_5 (\slash{\hskip -2mm}q+m_P )\gamma^{\nu}
\slash{\hskip -3mm}P_l \gamma_{\mu} (\slash{\hskip -2mm}p +m_B)
\gamma_5 \gamma_{\nu} \right]}{k^2 P_l^2 }
+\frac{Tr \left[\gamma_5 (\slash{\hskip -2mm}q+m_P )\gamma_{\mu}
(\slash{\hskip -3mm}P_b +m_b )\gamma^{\nu}(\slash{\hskip -2mm}p +m_B)
\gamma_5 \gamma_{\nu} \right]}{k^2 (P_b^2 -m_b^2 )}
\right\}~.\nonumber
\end{eqnarray}
In order to get quantitative estimation, we take the wave functions as
\cite{brod,carlson} 
\begin{eqnarray}
\phi_B (x)=\frac{f_B}{2\sqrt{6}}\delta(x-\epsilon_B )~,~~~~~~~~~~~
\phi_P (y)=\sqrt{\frac{3}{2}}f_P y(1-y) ~.
\end{eqnarray}
(here $\epsilon_B$ is the peaking position of the B-meson wave function.
Typically $\langle\epsilon_B \rangle \sim \frac{m_B-m_b}{m_B}$)
  We get
 \begin{equation}
 \langle P \vert \bar q_l \gamma_{\mu}b \vert B \rangle ~_{PQCD}
 =F_{+}^{B\rightarrow P}(Q^2)(p+q)_{\mu}
+F_{-}^{B\rightarrow P}(Q^2)(p-q)_{\mu}
 \end{equation}
 where
 \begin{eqnarray}
 F_{+}^{B\rightarrow P}(Q^2)=&&-\frac{8\pi \alpha_s}{3}f_Pf_B \left \{
 -\frac{m_Pm_B}{\epsilon_B^2m_B^4} \right. \nonumber\\
 &&\left. -\int dy y \frac{m_b(m_P-2m_B)+y(m_B^2-2m_Pm_B)}
 {\epsilon_Bm_B^2(ym_B^2-m_b^2)} \right \}, \\
 F_{-}^{B\rightarrow P}(Q^2)=&&-\frac{8\pi \alpha_s}{3}f_Pf_B \left \{
 -\frac{m_B(\epsilon_B-m_P)}{\epsilon_B^2m_B^4} \right. \nonumber\\
 &&\left. -\int dy y \frac{2m_b-4m_P-y(m_B-2m_P)}
 {\epsilon_Bm_B(ym_B^2-m_b^2)} \right \}.
 \end{eqnarray}
 Here, $Q^2=(p-q)^2$.So we can obtain the matrix element $M_{q_1q_2q_3}^{P'P}$.
 We can also get the matrix element $\langle P\vert \bar{q}_l b\vert B \rangle$
 as
\begin{eqnarray}
\langle P \vert \bar s b \vert B\rangle ~_{PQCD}&=&-\frac{2}{3}g^2
\int dxdy \phi_B (x)\phi_P (y)
\left\{
\frac{Tr \left[\gamma_5 (\slash{\hskip -2mm}q+m_P )\gamma^{\nu}
\slash{\hskip -3mm}P_l (\slash{\hskip -2mm}p +m_B)
\gamma_5 \gamma_{\nu} \right]}{k^2 P_l^2 }\right.\nonumber\\
& &\left.+\frac{Tr \left[\gamma_5 (\slash{\hskip -2mm}q+m_P )
(\slash{\hskip -3mm}P_b +m_b )\gamma^{\nu}(\slash{\hskip -3mm}P +m_B)
\gamma_5 \gamma_{\nu} \right]}{k^2 (P_b^2 -m_b^2 )}
\right\} \nonumber\\
&=&-\frac{8\pi}{3}\alpha_s f_B f_P 
\left\{
\frac{-2m_P(1-2\epsilon_B)+\epsilon_Bm_B)m_B+m_P^2}
{\epsilon_B^2 m_B^3 }\right.\nonumber\\
&~&\left.- \int dy y\frac{m_B(m_P+m_b)-2m_B^2-4m^bm_P+ym_Pm_B}
{\epsilon_B m_B (ym_B^2-m_b^2 )}
\right\}.
\end{eqnarray}

 In the literature\cite{zhu}, as an example the authors calculated the
numerical results of the matrix element $\langle K^{-} \vert \bar{s}
\gamma_{\mu} b \vert B^{-} \rangle$ by above framework, where they applied
$\alpha_s \simeq 0.38$, $f_B=200MeV$ and $f_K=160MeV$. One can find 
their results are sensitive to the values of parameters $\epsilon_B$ and 
$m_b$, and seem small compared with the $BSW$ result. We also compute
the matrix element $\langle \pi^{-} \vert \bar{s} \gamma_{\mu} b 
\vert B^{-} \rangle$ and list the numerical results in Table 1.
Where we take $\alpha_s=0.38$, $f_B=0.2GeV$ and $f_{\pi}=0.13GeV$. 

  We can see that the results are very sensitive to the values of parameter 
$\epsilon_B$ and $m_b$, and smaller than the BSW result which is about $0.29$
\cite{BSW2}. As mentioned in the Ref.\cite{zhu}, the PQCD results are 
comparatively small in many cases. 
  
  But the ratio 
\begin{equation}
{\cal{R}}=\frac{X_{q_1q_2q_3~PQCD}^{PP'}}{M_{q_1q_2q_3~PQCD}^{PP'}}
\end{equation}
is insensitive to the parameters $\epsilon_B$ and $m_b$. So it is more reliable
because of cancelation of the main uncertainties. We list our computation 
in the Table 2.

  The ratio by using equation of motion is
\begin{equation}
{\cal{R}}=
\frac{X_{sdd}^{\bar{K}^0 \pi^{-}}}{M_{q_1q_2q_3}^{\bar{K}^0 \pi^{-}}}
=\frac{m_{\bar{K}^0}^2}{(m_s+m_d)(m_d-m_b)}\simeq -0.30, 
\end{equation}  
and it is about one order of magnitude larger than the PQCD estimation.
As mentioned in our introduction, the matrix elements of $(S+P)(S-P)$
four quarks operator are very important in some decay modes of $B$ mesons, like
$B$ to $\eta'$ and other mesons. 
So it is necessary to recalculate the branching ratios and $CP$ asymmetries
for 2-body charmless $B$ decays by using the PQCD method instead of the
equation of motion. 

\section{branching ratios and CP asymmetries}
In the $B$ rest frame, the two body decay width is
\begin{equation}
\Gamma(B\rightarrow PP')=\frac {1}{8\pi}
\vert \langle PP' \vert H_{eff} \vert B \rangle \vert ^2
\frac{\vert p \vert}{m_B^2}, 
\end{equation}
where
\begin{equation}
\vert p \vert =\frac
{\left[(m_B^2-(m_P+m_{P'})^2)(m_B^2-(m_P-m_{P'})^2)\right]^{\frac {1}{2}}}
{2m_B}
\end{equation}
is the magnitude of the momentum of the particle $P$ or $P'$. The 
corresponding branching ratio is given by
\begin{equation}
{\cal{B}}_{BR}(B\rightarrow PP')
=\frac{\Gamma(B\rightarrow PP')}{\Gamma_{tot}}.
\end{equation}

The direct $CP$ asymmetry ${\cal{A}}_{CP}$ for $B$ meson decays into $PP^{'}$
is defined as
\begin{equation}
{\cal{A}}_{CP}=\frac {\Gamma(B\rightarrow PP')
-\Gamma(\bar{B}\rightarrow \bar{P}\bar{P}')}
{\Gamma(B\rightarrow PP')+\Gamma(\bar{B}\rightarrow \bar{P}\bar{P}')}.
\end{equation}
 
In our numerical calculation, we use the Wolfstein parameterization for 
the $CKM$ matrix
\begin{equation}
V_{CKM}=\left[
\begin{array}{ccc}
1-\frac{\lambda^2}{2}&\lambda& \lambda^3 A (\rho-i \eta) \\
-\lambda  &1-\frac{\lambda^2}{2}& \lambda^2 A \\
\lambda^3 A (1-\rho-i \eta) &-\lambda^2 A& 1
\end{array} \right]+{\cal{O}}(\lambda^4)
\end{equation}
and we take\cite{italy}
\begin{equation}
A=0.823\pm 0.033,~~~~ \lambda=0.220,~~~~\rho=0.160,~~~~\eta=0.336. 
\end{equation}
Otherwise we take all parameters such as meson decay constants and form 
factors needed in our calculation as follows\cite{yangmz,zheng}:
$f_{\pi}=0.13 GeV$, $f_K=0.160GeV$, $f_{\eta^{'}}^u=f_{\eta^{'}}^d=0.049GeV$,
$f_{\eta}^u=f_{\eta}^d=0.092GeV$, $f_{\eta^{'}}^s=0.12GeV$,
$f_{\eta}^s=-0.105GeV$, $f_{\eta{'}}^c=-0.0063GeV$, $f_{\eta}^c=-0.0024GeV$,
and $F_{+}^{ B_u^{-} \rightarrow \pi^{-} }(0)=0.29$,
$F_{+}^{ B_u^{-} \rightarrow K^{-} }(0)=0.32$,
$F_{+}^{ B_u^{-} \rightarrow \eta^{'} }(0)=\frac{0.254}{\sqrt{6}} $,
$F_{+}^{\bar{B}_s^{-}\rightarrow\eta^{'}}(0)=\frac{2\times 0.282}{\sqrt{6}}$,
$F_{+}^{B_u^{-}\rightarrow\eta}(0)=\frac{0.307}{\sqrt{3}}$,
$F_{+}^{\bar{B}_s^{-}\rightarrow\eta}(0)=-\frac{0.335}{\sqrt{3}}$.
Here we apply the flavor wave functions of $\eta^{'}$ and $\eta$ as\cite{zheng}
\begin{equation}
\left\{
\begin{array}{l}
\vert \eta^{'}\rangle=\frac{\vert u\bar{u}\rangle+\vert d\bar{d}\rangle
+2\vert s\bar{s}\rangle}{\sqrt{6}}\\
\\
\vert \eta\rangle=\frac{\vert u\bar{u}\rangle+\vert d\bar{d}\rangle
-\vert s\bar{s}\rangle}{\sqrt{3}}.
\end{array}
\right.
\end{equation}
  
  We give the numerical results of the branching ratios and $CP$ asymmetries
for $B$ charmless decays in Table 3,4,5. As a comparison, the results by using
the equation of motion are also listed in the tables where we take
$m_u=5MeV$, $m_d=10MeV$, $m_s=150MeV$ and $m_b=5.0GeV$. 
In the calculation, we have neglected the contributions of 
W-annihilation, W-exchange and space-like penguin diagrams.
 
  From the tables, we can see the following features:
  
 (i) For most of charmless B decays, the contributions of penguin diagrams
are important. 

 (ii) Comparing the results of the PQCD method with those by using
equation of motion, one can find large difference between them.
In the modes of $B\rightarrow \pi K$ and $B\rightarrow K K$, 
the branching ratios predicted by using equation of motion 
is larger than those of the PQCD  method by about a factor 2.
While final state involving $\eta^{'}$, the factor would be more large. 
Obviously, $CP$ asymmetries are also affected by these difference. 
In our computation, we find that the ratio of $X_{q_1q_2q_3}^{PP'}$ to
$M_{q_1q_2q_3}^{PP'}$ predicted by PQCD method is not of $m_P^2$ dependence 
like the estimation by use of equation of motion. So while $m_P$ is large,
the distinguishes between two method are more obvious.

 (iii)In many decay modes, the branching ratios are sensitive to the
color parameter $N_c^{eff}$, such as $\bar{B}_d^0\rightarrow \pi^0\pi^0$,
$\eta\eta(\eta^{'}\eta^{'})$ and $\eta\eta^{'}$ . Otherwise,
the value of $N_c^{eff}$ affects $CP$ asymmetries more largely than
branching ratios in some modes, for example, $\bar{B}_s^0\rightarrow 
K^0\eta^{'}$, which $CP$ asymmetry rangs from $60.6\%$ to $-50.1\%$ for
$N_c^{eff}$ ranging form $2$ to $\infty$. It is because that $a_i$ are 
sensitive to $N_c^{eff}$ which gives the different strong phases.

 (iv)Our results are smaller than those in some literatures\cite{yangmz,zheng}.
In some decay modes like $B\rightarrow K\eta'$, our results are one order of
magnitude smaller than the results of the experiments\cite{CLEO}. Because
we did not consider the contributions of other mechanisms, such as 
$b\rightarrow sg^{*}\rightarrow s\eta^{'}$ via $QCD$ anormaly\cite{anormaly},
$b\rightarrow sgg \rightarrow s\eta{'}$\cite{yyd} $\it{etc.}$.
In the Ref.\cite{zhu,korea}, the authors gave the numerical results
involved the contributions of the new mechanisms, which fit the experiments
very well. 

\section{concluding remarks}
In this paper, we recalculate the decays of $B$ to two charmless pseudoscalar
mesons with conventional method (the standard effective weak Hamiltonian and
the $BSW$ model). Instead of using equations of motion, we use an 
alternative method to estimate the hadronic matrix elements $(S+P)(S-P)$
and obtain comparatively smaller results. In some modes, which are penguin
dominant, such as $B\rightarrow \pi K$, the branching ratios that we
predicted seem to be a little bit smaller than the lower limits of the
experiments of $CLEO$\cite{CLEO}. But they are derived in the factorization
approach, many mechanisms are not considered in this work such as final state
interactions. Especially in the modes of $B\rightarrow \pi K$ or $KK$, FSI could
yield dominant contribution to the decay width\cite{Ali}. So more uncertainties
in non-leptonic charmless B decays need us to study in the future.

\section{acknowledgement}
This work is supported in part by National Science Foundation of China and 
State Commission of Science and Technology of China.


\newpage

\narrowtext
\tighten

\begin{table}
\vspace*{5cm}
\begin{tabular}{c|cccc}
$F^{B_u^{-}\rightarrow\pi^{-}}_{+}$ & $\epsilon_B=0.05$ & $\epsilon_B=0.06$ &
$\epsilon_B=0.07$ & $\epsilon_B=0.08$\\\hline
$mb=5.0GeV$ & $0.21$ & $0.17$ & $0.14$ & $0.12$ \\\hline
$mb=4.9GeV$ & $0.26$ & $0.21$ & $0.18$ & $0.15$ \\\hline
$mb=4.8GeV$ & $0.19$ & $0.15$ & $0.13$ & $0.11$ \\\hline
\end{tabular}
\vspace*{0.5cm}
\caption{The PQCD estimations about the element $\langle \pi^{-} \vert 
\bar{s} \gamma_{\mu} b \vert B^{-} \rangle$.}
\end{table}

\begin{table}
\vspace*{5cm}
\begin{tabular}{c|cccc}
${\cal{R}}=\frac{X_{sdd~PQCD}^{\bar{K}^0 \pi^{-}}}
{M_{sdd~PQCD}^{\bar{K}^0 \pi^{-}}}$ 
& $\epsilon_B=0.05$ & $\epsilon_B=0.06$ &
$\epsilon_B=0.07$ & $\epsilon_B=0.08$\\\hline
$mb=5.0GeV$ & $-0.025$ & $-0.025$ & $-0.026$ & $-0.026$ \\\hline
$mb=4.9GeV$ & $-0.025$ & $-0.026$ & $-0.026$ & $-0.026$ \\\hline
$mb=4.8GeV$ & $-0.026$ & $-0.026$ & $-0.027$ & $-0.027$ \\\hline
\end{tabular}
\vspace*{0.5cm}
\caption{The PQCD estimations about the ratio of the matrix element
$X_{sdd~PQCD}^{\bar{K}^0 \pi^{-}}$ to $M_{sdd~PQCD}^{\bar{K}^0 \pi^{-}}$.}
\vspace*{0.5cm}
\end{table} 

\newpage

\begin{table}
\vspace*{1.5cm}
\caption{Branching ratio in $10^{-5}$,and $CP$ asymmetries in $\%$.
'QCD' and 'EW' present the QCD penguin and EW penguin effects
respectively, and 'DIRAC' presents the results with the equation of 
motion.}
\vspace*{0.5cm}
\begin{tabular}{lccccccc}
$N_c^{eff}=2$ & \multicolumn{4}{c}{Branching~Ratio} &
\multicolumn{3}{c}{CP~Asymmetry} \\
Decay Mode&TR&QCD&EW&DIRAC&QCD&EW&DIRAC\\\hline
 $B_u^{-}\rightarrow\pi^0 \pi^{-}$& $0.53$ & $0.53$ & $0.53$ & $0.54$ 
 & $0.03$ & $1.9$ & $3.5$  \\
 $B_u^{-}\rightarrow \pi^{-}\eta^{'}$ & $0.14$ & $0.19$ & $0.19$ & $1.54$ 
 & $16.4$ & $0.05$ & $0.2$ \\
 $B_u^{-}\rightarrow \pi^{-}\eta$ & $0.42$ & $0.57$ & $0.57$ & $1.48$ &
  $16.0$ & $0.1$ & $0.6$ \\
 $B_u^{-}\rightarrow K^0 K^{-}$ & $~$ & $0.032$ & $0.032$ & $0.065$ 
 & $12.9$ & $13.3$ & $12.3$ \\ 
  $B_u^{-}\rightarrow \pi^0 K^{-}$ & $0.038$ & $0.090$ & $0.19$ & $0.405$
  & $-39.0$ & $-23.3$ & $-14.8$ \\  
 $B_u^{-}\rightarrow K^{-}\eta^{'}$ & $0.17$ & $0.42$ & $0.38$ & $1.38$
 & $-11.5$ & $-16.7$ & $-7.8$ \\
 $B_u^{-}\rightarrow K^{-}\eta$ & $0.024$ & $0.063$ & $0.038$ & $0.025$ 
 & $17.8$ & $-7.4$ & $-10.4$ \\ 
 $B_u^{-}\rightarrow \bar{K}^0 \pi^{-}$ & $~$ & $0.35$ & $0.33$ & $0.72$
 & $-0.3$ & $-0.3$ & $-0.3$ \\
 $\bar{B}_d^{0}\rightarrow \pi^{+} \pi^{-}$ & $0.64$ & $0.76$ & $0.83$ 
 & $0.77$ & $13.2$ & $13.2$ & $18.1$ \\ 
 $\bar{B}_d^{0}\rightarrow \pi^{0} \pi^{0}$ & $0.012$ & $7.9\times 10^{-3}$
 & $6.8\times 10^{-3}$ & $6.94\times 10^{-3}$ & $-42.7$ & $-46.8$ & $-51.5$\\
 $\bar{B}_d^0\rightarrow \pi^{0} \eta^{'}$ & $1.6\times 10^{-3}$
 &$7.0\times 10^{-3}$ & $5.0\times 10^{-3}$ & $5.0\times 10^{-3}$
 &$-27.8$ & $-36.0$ & $-36.0$  \\
 $\bar{B}_d^0\rightarrow \pi^{0} \eta$ & $3.1\times 10^{-4}$
 &$0.019$ &$0.019$ &$0.34$ & $0.8$ & $0.8$ & $8.6$ \\
 $\bar{B}_d^0\rightarrow \eta^{'} \eta^{'}$ & $1.1\times 10^{-3}$    
 &$1.2\times 10^{-3}$ &$1.1\times 10^{-3}$ & $8.63\times 10^{-3}$
 & $7.5$ & $5.6$ & $12.2$ \\  
 $\bar{B}_d^0\rightarrow \eta \eta$ & $9.4\times 10^{-3}$
 &$7.6\times 10^{-3}$ &$8.1\times 10^{-3}$ & $8.47\times10^{-3}$ 
 & $2.2$ & $4.8$ & $35.5$ \\
 $\bar{B}_d^0\rightarrow \eta \eta^{'}$ & $0.056$        
 & $0.058$ &$0.056$ & $0.094$ & $9.3$ & $9.3$ & $95.7$ \\
 $\bar{B}_d^0\rightarrow K^0 \bar{K}^{0}$ & $~$            
 & $0.042$ &$0.041$ & $0.065$ & $12.8$ & $12.6$ & $12.3$ \\
 $\bar{B}_d^0\rightarrow \pi^{+} K^{-}$ & $0.048$   
 & $0.16$ &$0.23$ & $0.46$ & $-31.7$ & $-28.5$ & $-18.2$ \\
 $\bar{B}_d^0\rightarrow \pi^{0} \bar{K}^{0}$ & $1.3\times 10^{-3}$
 & $0.22$ & $0.14$ & $0.38$ & $5.3$ & $8.7$ & $3.9$  \\
 $\bar{B}_d^0\rightarrow \bar{K}^{0}\eta^{'}$ & $0.017$ 
 &$0.42$ &$0.36$ & $1.47$ & $-2.9$ & $-10.8$ & $-5.4$ \\
 $\bar{B}_d^0\rightarrow \bar{K}^{0}\eta$ & $2.4\times 10^{-3}$
 &$0.013$ &$3.6\times 10^{-3}$ & $2.8\times10^{-3}$ 
 & $13.7$ & $-45.2$ & $2.3$ \\
 $\bar{B}_s^0\rightarrow \pi^{-} K^{+}$ & $0.68$
 &$0.81$ &$0.81$ & $0.89$ & $13.1$ & $13.1$ & $18.1$ \\
 $\bar{B}_s^0\rightarrow \pi^{0} K^{0}$ & $0.020$
 &$0.012$ &$0.012$ & $0.012$ & $-40.9$ & $-44.7$ & $-45.8$ \\
 $\bar{B}_s^0\rightarrow K^0 \eta^{'}$ & $8.2\times 10^{-3}$            
 &$0.063$ &$0.060$ & $1.16$ & $62.2$ & $60.6$ & $19.5$ \\ 
 $\bar{B}_s^0\rightarrow K^0 \eta$ & $0.022$            
 &$0.019$ &$0.019$ & $0.34$ & $-8.7$ & $-4.9$ & $34.9$ \\ 
 $\bar{B}_s^0\rightarrow K^0 \bar{K}^{0}$ & $~$            
 &$0.35$ &$0.33$ & $0.75$ & $-0.3$ & $-0.3$ & $-0.3$ \\
 $\bar{B}_s^0\rightarrow K^{-} K^{+}$ & $~$            
 &$0.31$ &$0.35$ & $0.75$ & $-0.3$ & $-0.3$ & $-0.3$ \\ 
 $\bar{B}_s^0\rightarrow \pi^{0} \eta^{'}$ & $6.1\times 10^{-4}$
 &$6.1\times 10^{-4}$ & $4.3\times 10^{-3}$ 
 & $4.3\times10^{-3}$ & $0$ & $0$ & $0$ \\
 $\bar{B}_s^0\rightarrow \pi^{0} \eta$ & $4.6\times 10^{-4}$
 &$4.6\times 10^{-4}$ & $3.3\times 10^{-3}$ 
 & $3.3\times10^{-3}$ & $0$ & $0$ & $0$ \\
 $\bar{B}_s^0\rightarrow \eta^{'} \eta^{'}$ & $7.4\times 10^{-3}$    
 & $0.11$ &$0.11$ & $0.41$ & $-3.7$ & $-3.8$ & $-1.7$ \\   
 $\bar{B}_s^0\rightarrow \eta \eta$ & $8.2\times 10^{-4}$    
 & $0.078$ &$0.078$ & $0.11$ & $4.8$ & $4.6$ & $3.2$ \\ 
 $\bar{B}_s^0\rightarrow \eta \eta^{'}$ & $3.1\times 10^{-4}$    
 & $0.33$ &$0.33$ & $0.82$ & $1.4$ & $1.2$ & $0.4$ \\
 \end{tabular}
\end{table}

\newpage

\narrowtext
\tighten

\begin{table}
\vspace*{1.5cm}
\caption{Branching ratio in $10^{-5}$,and $CP$ asymmetries in $\%$.
'QCD' and 'EW' present the QCD penguin and EW penguin effects
respectively, and 'DIRAC' presents the results with the equation of  
motion.}
\vspace*{0.5cm}
\begin{tabular}{lccccccc}
$N_c^{eff}=3$ & \multicolumn{4}{c}{Branching~Ratio} &
\multicolumn{3}{c}{CP~Asymmetry} \\
Decay Mode&TR&QCD&EW&DIRAC&QCD&EW&DIRAC \\\hline
 $B_u^{-}\rightarrow\pi^0 \pi^{-}$& $0.42$ & $0.42$ & $0.42$ & $0.43$
 & $0.04$ & $1.9$ & $3.8$ \\
 $B_u^{-}\rightarrow \pi^{-}\eta^{'}$ & $0.10$ & $0.15$ & $0.15$ & $1.56$ 
 & $18.9$ & $0.04$ & $0.2$ \\
 $B_u^{-}\rightarrow \pi^{-}\eta$ & $0.31$ & $0.46$ & $0.46$ & $1.39$
 & $19.2$ & $0.2$ & $0.7$ \\
 $B_u^{-}\rightarrow K^0 K^{-}$ & $~$ & $0.037$ & $0.036$ & $0.075$ 
 &  $12.8$ & $12.9$ & $12.1$ \\
 $B_u^{-}\rightarrow \pi^0 K^{-}$ & $0.031$ & $0.12$ & $0.22$ & $0.46$
 &  $-32.3$ & $-20.4$ & $-12.9$  \\
 $B_u^{-}\rightarrow K^{-}\eta^{'}$ & $5.9\times 10^{-3}$ 
 & $0.42$ & $0.38$ & $1.44$
 & $-11.0$ & $-15.9$ & $-7.2$ \\ 
 $B_u^{-}\rightarrow K^{-}\eta$ & $0.021$ & $0.067$ & $0.039$ & $0.021$ 
 & $15.4$ & $-11.1$ & $19.8$ \\ 
 $B_u^{-}\rightarrow \bar{K}^0 \pi^{-}$ & $~$ & $0.41$ & $0.40$ & $0.84$
 &  $-0.3$ & $-0.3$ & $-0.3$ \\ 
 $\bar{B}_d^{0}\rightarrow \pi^{+} \pi^{-}$ & $0.71$ & $0.85$ & $0.85$ 
 & $0.92$ & $13.3$ & $13.3$ & $18.3$ \\
 $\bar{B}_d^{0}\rightarrow \pi^0 \pi^0$ & $8.8\times 10^{-4}$ & 
 $2.8\times 10^{-3}$ & $1.4\times 10^{-3}$ & $3.6\times 10^{-3}$
 & $-31.8$ & $-50.3$ & $-30.3$ \\
 $\bar{B}_d^{0}\rightarrow \pi^0 \eta^{'}$ & $1.2\times 10^{-4}$ &
 $9.3\times 10^{-3}$ & $6.9\times 10^{-3}$ & $6.9\times 10^{-3}$
 & $3.2$ & $2.3$ & $2.3$ \\
 $\bar{B}_d^{0}\rightarrow \pi^0 \eta$ & $2.2\times 10^{-5}$ &
 $0.023$ & $0.023$ & $0.38$ 
 & $10.4$ & $10.8$ & $10.8$ \\
 $\bar{B}_d^0\rightarrow \eta^{'} \eta^{'}$ & $8.1\times 10^{-5}$
 &$1.1\times 10^{-4}$ &$7.8\times 10^{-5}$ & $0.013$ 
 & $28.3$ & $0.3$ & $1.1$ \\
 $\bar{B}_d^0\rightarrow \eta \eta$ & $6.8\times 10^{-4}$        
 &$4.1\times 10^{-4}$ &$7.9\times 10^{-4}$ & $0.017$ 
 & $13.2$ & $61.7$ & $12.1$ \\
 $\bar{B}_d^0\rightarrow \eta \eta^{'}$ & $4.0\times 10^{-3}$            
 & $4.7\times 10^{-3}$ &$4.2 \times 10^{-3}$ & $0.073$
 & $42.2$ & $42.8$ & $21.0$ \\
 $\bar{B}_d^0\rightarrow K^0 \bar{K}^{0}$ & $~$   
 & $0.049$ & $0.048$ & $0.076$ 
 & $12.4$ & $12.5$ & $12.1$ \\
 $\bar{B}_d^0\rightarrow \pi^{+} K^{-}$ & $0.053$ & $0.23$ & $0.25$ 
 & $0.50$ & $-30.4$ & $-29.1$ & $-18.7$ \\
 $\bar{B}_d^0\rightarrow \pi^0 \bar{K}^{0}$ & $9.7\times 10^{-5}$
 &$0.24$ &$0.15$ & $0.42$ 
 & $1.1$ & $1.8$ & $0.7$ \\
 $\bar{B}_d^0\rightarrow \bar{K}^{0}\eta^{'}$ & $1.2\times 10^{-3}$           
 &$0.43$ &$0.38$ & $1.56$ 
 & $-1.1$ & $-7.2$ & $-3.7$ \\
 $\bar{B}_d^0\rightarrow \bar{K}^{0}\eta$ & $1.7\times 10^{-4}$
 & $0.015$ & $2.8\times 10^{-3}$ & $4.9\times 10^{-3}$
 & $2.8$ & $-67.9$ & $-78.5$ \\
 $\bar{B}_s^0\rightarrow \pi^{-} K^{+}$ & $0.75$
 & $0.90$ & $0.90$ & $0.98$
 & $13.3$ & $13.3$ & $18.3$ \\
 $\bar{B}_s^0\rightarrow \pi^{0} K^{0}$ & $1.4\times 10^{-3}$
 & $4.3\times 10^{-3}$ & $2.0\times 10^{-3}$ & $2.2\times 10^{-3}$
 & $-32.6$ & $-52.3$ & $-49.6$ \\
 $\bar{B}_s^0\rightarrow K^0 \eta^{'}$ & $5.9\times 10^{-4}$            
 &$0.043$ &$0.041$ & $1.2$ 
 & $30.9$ & $29.0$ & $13.3$ \\
 $\bar{B}_s^0\rightarrow K^0 \eta$ & $1.6\times 10^{-3}$            
 &$1.3\times 10^{-3}$ &$9.3\times 10^{-4}$ & $0.29$
 & $-32.6$ & $-38.1$ & $19.8$ \\ 
 $\bar{B}_s^0\rightarrow K^0 \bar{K}^{0}$ & $~$            
 &$0.41$ &$0.40$ & $0.88$
 & $-0.3$ & $-0.3$ & $-0.3$ \\ 
 $\bar{B}_s^0\rightarrow K^{-} K^{+}$ & $~$            
 &$0.37$ &$0.39$ & $0.82$ 
 & $-0.3$ & $-0.3$ & $-0.3$ \\
 $\bar{B}_s^0\rightarrow \pi^{0} \eta^{'}$ & $4.4\times 10^{-5}$
 &$4.4\times 10^{-5}$ & $5.3\times 10^{-3}$ & $5.3\times 10^{-3}$
 & $0$ & $0$ & $0$ \\
 $\bar{B}_s^0\rightarrow \pi^{0} \eta$ & $3.3\times 10^{-5}$
 &$3.3\times 10^{-5}$ & $4.0\times 10^{-3}$ & $4.0\times 10^{-3}$
 & $0$ & $0$ & $0$ \\ 
 $\bar{B}_s^0\rightarrow \eta^{'} \eta^{'}$ & $5.3\times 10^{-4}$    
 &$0.11$ &$0.11$ & $0.42$ 
 & $-1.4$ & $-1.5$ & $-0.8$ \\   
 $\bar{B}_s^0\rightarrow \eta \eta$ & $5.9\times 10^{-5}$    
 &$0.092$ &$0.092$ & $0.13$
 & $0.9$ & $0.7$ & $0.5$ \\
 $\bar{B}_s^0\rightarrow \eta \eta^{'}$ & $2.2\times 10^{-5}$    
 &$0.39$ &$0.39$ & $0.93$ & $0.06$ & $-0.1$ & $-0.2$ \\          
\end{tabular}
\end{table}

\newpage

\narrowtext
\tighten

\begin{table}
\vspace*{1.5cm}
\caption{Branching ratios in unit of $10^{-5}$,and $CP$ 
asymmetry in unit of $\%$. 'QCD'
and 'EW' present the QCD penguin and EW penguin effects respectively,
and 'DIRAC' presents the results with the equation of  
motion.}
\vspace*{0.5cm}
\begin{tabular}{lccccccc}
$N_c^{eff}=\infty$ & \multicolumn{4}{c}{Branching~Ratio} &
\multicolumn{3}{c}{CP~Asymmetry} \\
Decay Mode&TR&QCD&EW&DIRAC&QCD&EW&DIRAC \\\hline
 $B_u^{-}\rightarrow\pi^0 \pi^{-}$& $0.23$ & $0.23$ & $0.24$ & $0.25$ 
 & $0.05$ & $1.9$ & $4.7$ \\
 $B_u^{-}\rightarrow \pi^{-}\eta^{'}$ & $0.045$ & $0.082$ & $0.082$ & $1.60$
 & $25.2$ & $0.2$ & $0.2$ \\
 $B_u^{-}\rightarrow \pi^{-}\eta$ & $0.14$ & $0.27$ & $0.27$ & $1.25$
 & $29.5$ & $0.5$ & $0.9$ \\
 $B_u^{-}\rightarrow K^0 K^{-}$ & $~$ & $0.048$ & $0.049$ & $0.10$ 
 & $12.5$ & $12.3$ & $11.9$ \\ 
 $B_u^{-}\rightarrow \pi^0 K^{-}$ & $0.019$ & $0.18$ & $0.28$ & $0.58$
 & $-21.1$ & $-14.9$ & $-9.5$ \\
 $B_u^{-}\rightarrow K^{-}\eta^{'}$ & $0.040$ & $0.42$ & $0.38$ & $1.56$
 & $-9.0$ & $-13.6$ & $-6.0$ \\ 
 $B_u^{-}\rightarrow K^{-}\eta$ & $0.022$ & $0.079$ & $0.054$ & $0.024$
 & $10.0$ & $-8.8$ & $20.9$ \\
 $B_u^{-}\rightarrow \bar{K}^0 \pi^{-}$ & $~$ & $0.54$ & $0.56$ & $1.12$
 & $-0.3$ & $-0.3$ & $-0.3$ \\ 
 $\bar{B}_d^0\rightarrow \pi^{+} \pi^{-}$ & $0.86$ & $1.00$ & $1.00$ 
 & $1.12$ & $13.6$ & $13.6$ & $18.7$ \\
 $\bar{B}_d^0\rightarrow \pi^{0} \pi^{0}$ & $0.017$ &$0.034$ & $0.032$ 
 & $0.039$ & $39.1$ & $42.0$ & $46.4$ \\
 $\bar{B}_d^0\rightarrow \pi^{0} \eta^{'}$ & $2.3\times 10^{-3}$
 &$0.020$ & $0.016$ & $0.017$ 
 & $61.3$ & $78.0$ & $78.0$ \\ 
 $\bar{B}_d^0\rightarrow \pi^{0} \eta$ & $4.5\times 10^{-4}$
 &$0.034$ & $0.031$ & $0.47$ 
 & $26.1$ & $29.2$ & $14.7$ \\
 $\bar{B}_d^0\rightarrow \eta^{'} \eta^{'}$ & $1.6\times 10^{-3}$
 &$1.8\times 10^{-3}$ &$1.8\times 10^{-3}$ & $0.026$
 & $-6.3$ & $-7.3$ & $-6.5$ \\
 $\bar{B}_d^0\rightarrow \eta \eta$ & $0.013$        
 &$0.017$ &$0.018$ & $0.070$
 & $-1.3$ & $-0.6$ & $1.1$ \\
 $\bar{B}_d^0\rightarrow \eta \eta^{'}$ & $0.079$            
 &$0.080$ &$0.080$ & $0.21$
 & $-8.1$ & $-8.1$ & $-24.1$ \\
 $\bar{B}_d^0\rightarrow K^0 \bar{K}^{0}$ & $~$   
 &$0.063$ &$0.064$ & $0.099$
 & $12.3$ & $12.1$ & $11.9$ \\
 $\bar{B}_d^0\rightarrow \pi^{+} K^{-}$ & $0.065$
 &$0.31$ &$0.29$ & $0.58$
 & $-28.4$ & $-30.4$ & $-19.5$ \\
 $\bar{B}_d^0\rightarrow \pi^0 \bar{K}^{0}$ & $1.9\times 10^{-3}$
 & $0.27$ & $0.18$ & $0.52$
 & $-5.9$ & $-8.5$ & $-4.4$ \\
 $\bar{B}_d^0\rightarrow \bar{K}^{0}\eta^{'}$ & $0.024$           
 & $0.44$ &$0.42$ & $1.77$ 
 & $3.1$ & $0.5$ & $-0.3$ \\
 $\bar{B}_d^0\rightarrow \bar{K}^{0}\eta$ & $3.4\times 10^{-3}$
 & $0.027$ & $6.8\times 10^{-3}$ & $1.5\times 10^{-3}$
 & $-7.8$ & $-63.9$ & $-79.6$ \\
 $\bar{B}_s^0\rightarrow \pi^{-} K^{+}$ & $0.91$
 & $1.10$ & $1.10$ & $1.19$
 & $13.5$ & $13.6$ & $18.7$ \\
 $\bar{B}_s^0\rightarrow \pi^{0} K^{0}$ & $0.029$
 & $0.055$ & $0.052$ & $0.049$
 & $36.6$ & $39.2$ & $36.4$ \\ 
 $\bar{B}_s^0\rightarrow K^0 \eta^{'}$ & $0.012$            
 & $0.021$ & $0.020$ & $1.27$
 & $-48.6$ & $-50.1$ & $1.7$ \\
 $\bar{B}_s^0\rightarrow K^0 \eta$ & $0.031$            
 & $0.037$ & $0.035$ & $0.24$
 & $14.2$ & $9.2$ & $-16.4$ \\   
 $\bar{B}_s^0\rightarrow K^0 \bar{K}^{0}$ & $~$            
 & $0.54$ & $0.56$ & $1.17$
 & $-0.3$ & $-0.3$ & $-0.3$ \\ 
 $\bar{B}_s^0\rightarrow K^{-} K^{+}$ & $~$            
 & $0.48$ & $0.45$ & $0.96$ 
 & $-0.3$ & $-0.3$ & $-0.3$ \\
 $\bar{B}_s^0\rightarrow \pi^{0} \eta^{'}$ & $8.7\times 10^{-4}$
 & $8.7\times 10^{-4}$ & $9.3\times 10^{-4}$ & $9.3\times 10^{-4}$
 & $0$ & $0$ & $0$ \\ 
 $\bar{B}_s^0\rightarrow \pi^{0} \eta$ & $6.5\times 10^{-4}$
 & $6.5\times 10^{-4}$ & $7.0\times 10^{-3}$ & $7.0\times 10^{-3}$
 & $0$ & $0$ & $0$ \\ 
 $\bar{B}_s^0\rightarrow \eta^{'} \eta^{'}$ & $0.011$    
 & $0.10$ & $0.10$ & $0.43$
 & $5.1$ & $5.0$ & $1.4$ \\
 $\bar{B}_s^0\rightarrow \eta \eta$ & $1.2\times 10^{-3}$    
 &$0.12$ &$0.12$ & $0.17$
 & $-4.4$ & $-4.5$ & $-3.5$ \\
 $\bar{B}_s^0\rightarrow \eta \eta^{'}$ & $4.4\times 10^{-4}$    
 & $0.50$ & $0.50$ & $1.2$ 
 & $-1.9$ & $-2.0$ & $-1.2$ \\       
\end{tabular}
\end{table}
\newpage

\begin{figure}[tb]
\vspace*{2cm}
\hspace*{-6cm}
\centerline{\epsfig{figure=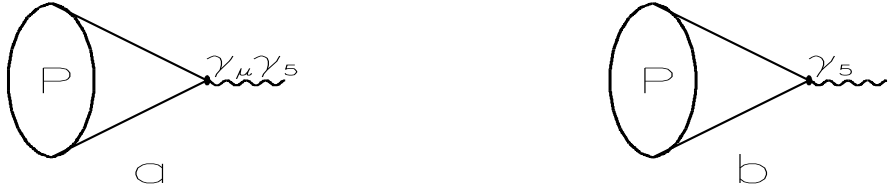,height=6cm,width=8cm,angle=0}}
\vspace*{-4cm}
\caption{\em Diagrams for the matrix elements $\langle P \mid \bar q_1
\gamma_{\mu} \gamma_5 q_2 \mid 0 \rangle$(fig 1a) and
$\langle P \mid \bar q_1 \gamma_5 q_2 \mid 0 \rangle$(fig 1b).}
\end{figure}

\begin{figure}[tb]
\vspace*{4.5cm}
\hspace*{-6cm}
\centerline{\epsfig{figure=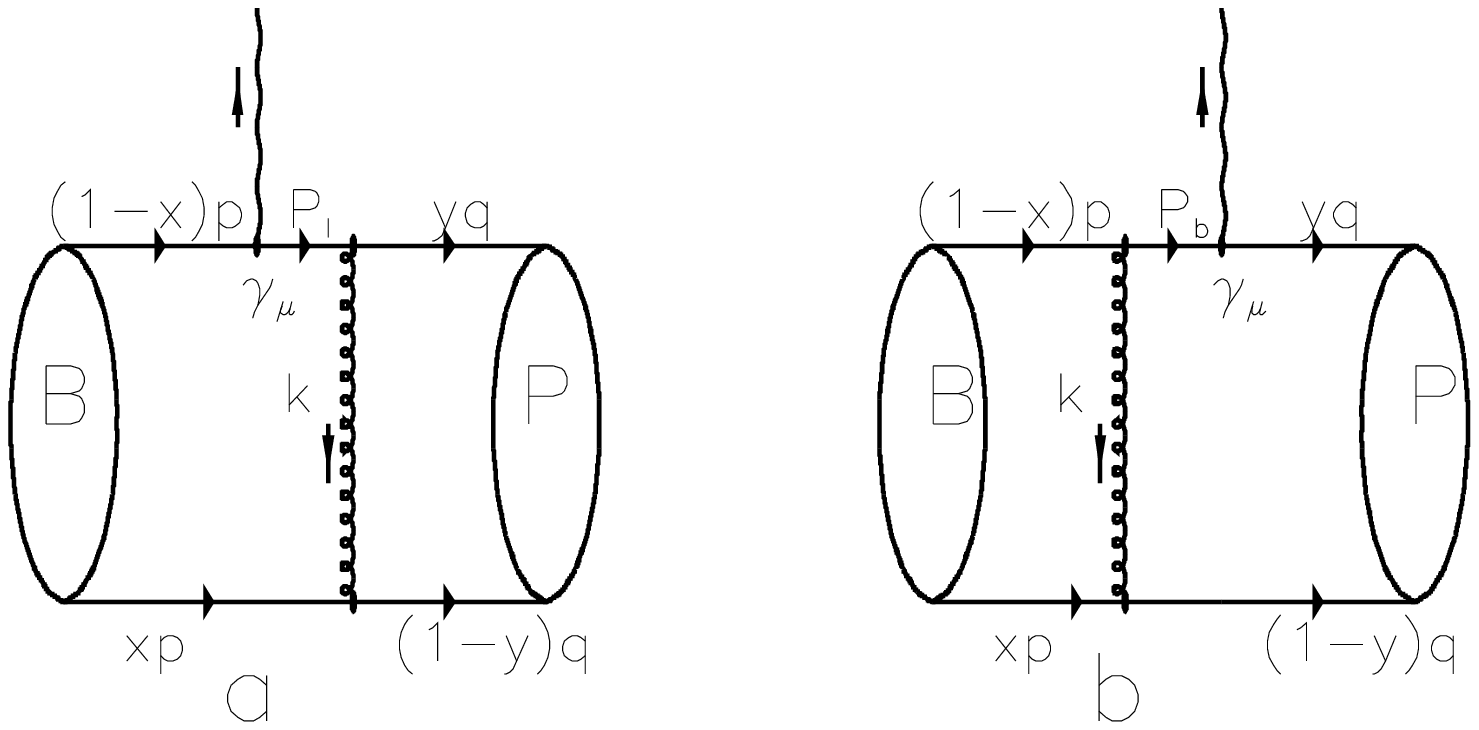,height=6cm,width=10cm,angle=0}}
\vspace*{-4cm}
\caption{\em Leading twist diagrams in $QCD$ for the 
matrix elements $\langle P\vert \bar q_l \gamma_{\mu} b\vert B \rangle$.} 
\end{figure}

\begin{figure}[tb]
\vspace*{4.5cm}
\hspace*{-6cm}
\centerline{\epsfig{figure=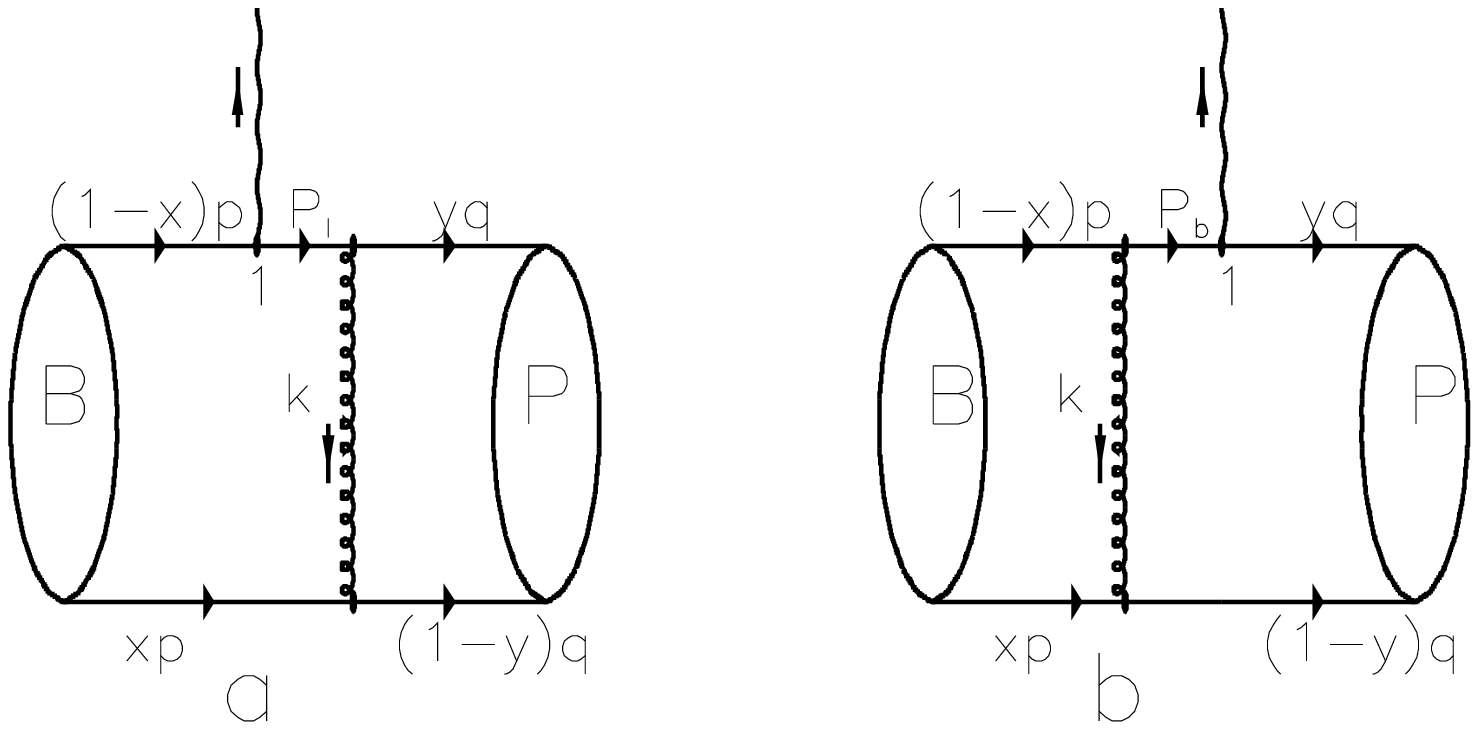,height=6cm,width=10cm,angle=0}}
\vspace*{-4cm}
\caption{\em Leading twist diagrams in $QCD$ for the 
matrix elements $\langle P\vert \bar{q}_l b\vert B \rangle$.} 
\end{figure}

\end{document}